\documentclass[slac_one]{revtex4}
\usepackage{graphicx}
\usepackage{fancyhdr}
\begin{document}
\def\thefootnote{\fnsymbol{footnote}}

\begin{titlepage}



\vspace*{33pt}

\title{
CALICE Si/W electromagnetic calorimeter prototype, first testbeam results 
}

\author{G. Mavromanolakis~\footnote[1]
{~email: {\tt gmavroma@hep.phy.cam.ac.uk} or {\tt gmavroma@mail.cern.ch}}}
\affiliation{Cavendish Laboratory, University of Cambridge, 
Cambridge CB3 0HE, UK}

\vspace{15pt} 
\begin{abstract}
{\small
A highly granular electromagnetic calorimeter prototype based on 
tungsten absorber and sampling units equipped with silicon pads as sensitive 
devices for signal collection is under construction. The full prototype 
will have in total 30 layers and be read out by about 10000 Si cells 
of 1~$\times$~1~cm$^2$.
A first module consisting of 14 layers and depth of 7.2~$X_0$ 
at normal incidence, having in total 3024 channels of 1~cm$^2$, was 
tested recently with $e^-$ beam. We describe the prototype and discuss 
some preliminary testbeam results.
}
\end{abstract}

\maketitle

\vfill
\begin{center}
Proceedings of the International Linear Collider Workshop \\
LCWS 2005, Stanford University, 18-22 March 2005
\end{center}
\vfill

\end{titlepage}

\thispagestyle{empty}

\clearpage

\setcounter{page}{1}

\title{{\small{2005 International Linear Collider Workshop - Stanford,
U.S.A.}}\\ 
\vspace{12pt}
CALICE Si/W electromagnetic calorimeter prototype, first testbeam results} 

%

\author{G. Mavromanolakis~\footnote[1]
{~email: {\tt gmavroma@hep.phy.cam.ac.uk} or {\tt gmavroma@mail.cern.ch}}}
\affiliation{Cavendish Laboratory, University of Cambridge, 
Cambridge CB3 0HE, UK}

\begin{abstract}

A highly granular electromagnetic calorimeter prototype based on 
tungsten absorber and sampling units equipped with silicon pads as sensitive 
devices for signal collection is under construction. The full prototype 
will have in total 30 layers and be read out by about 10000 Si cells 
of 1~$\times$~1~cm$^2$.
A first module consisting of 14 layers and depth of 7.2~$X_0$ 
at normal incidence, having in total 3024 channels of 1~cm$^2$, was 
tested recently with $e^-$ beam. We describe the prototype and discuss 
some preliminary testbeam results.

\end{abstract}

\maketitle

\thispagestyle{fancy}


\section{GENERAL} 

An experiment at the $e^{+} e^{-}$ Future Linear Collider with 0.5 - 1~TeV 
center-of-mass energy range must be capable to perform high precision 
measurements in order to exploit the physics potential of the machine.
This fact sets strict requirements on performance of vertex, tracking 
and calorimetric detectors. The CALICE Collaboration~\cite{ref:calice} has 
been formed to conduct the research and development effort needed to 
bring initial conceptual designs for the calorimetry to a final proposal 
suitable for an experiment at the Future Linear Collider.

The main proposal is that both electromagnetic and hadronic  
calorimeters should be highly granular to allow very efficient pattern
recognition for excellent shower separation and particle identification within 
jets and subsequently to provide excellent jet reconstruction efficiency 
\cite{ref:calorimetry}. This concept, also known as 
``particle flow paradigm'', to be successful requires optimal interplay 
between hardware, {\em i.e.} granularity, and software, {\em i.e.} 
reconstruction algorithms.

CALICE plans include studies of both electromagnetic and hadronic calorimeter 
prototypes. The electromagnetic prototype is a sampling calorimeter 
with W absorber and Si pads as sensitive material. There are two main 
different hadronic prototype concepts under study both based on steel 
absorber. One has scintillator tiles and conventional analogue readout 
scheme, while the other is envisaged to be equipped with resistive plate 
chambers or GEMs and have digital readout \cite{ref:calorimetry}.

Combined and individual testbeam studies are planned that will allow us 
to debug technology/detector concept(s), to perform detector characterisation, 
to test the ``particle flow paradigm'' and the interplay between hardware 
and software, and also to test-validate-improve available simulation codes and
shower packages \cite{ref:gm}.

The Si/W ECAL prototype and first testbeam results are discussed here.

\section{Si/W ECAL PROTOTYPE}

\begin{figure*}[tbp]
\centering
\includegraphics[width=135mm]{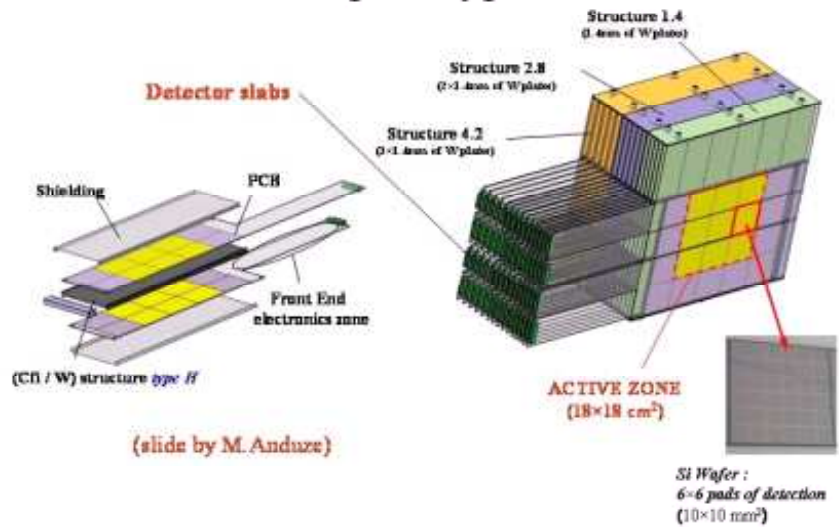}
\caption{Schematic of the full Si/W electromagnetic prototype.} 
\label{fig:EcalProto}
\end{figure*}

The Si/W electromagnetic prototype is a sampling calorimeter with high 
granularity, in both transverse and longitudinal direction. The full 
detector is longitudinally segmented into 30 layers of W interleaved 
with 0.5~mm thick Si pads as sensitive material. The W layers have 
varying thickness, the first 10 layers are 1.4~mm thick each, 2.8~mm in 
the next 10 and 4.2~mm in the final 10. The total depth of the 
detector is about 24~$X_0$ at normal incidence, and has an active face 
with an area of 18~$\times$~18~cm$^2$. It is read out in 
1~$\times$~1~cm$^2$ cells and in total there are 9720 channels arranged 
in 270 wafers each consisting of a 6~$\times$~6 cell matrix. 
A schematic of the full prototype is given in Figure~\ref{fig:EcalProto} 
which illustrates more clearly the actual construction. The layers are 
arranged in slabs with their supporting structure consisting of H-shaped W 
sheet wrapped with a carbon fiber layer. On each side of the W sheet of the 
slabs PCB's are put on which the Si wafers are glued. Then each slab is 
shielded with aluminum foil. This construction introduces odd/even layer 
asymmetry, which contributes to difference in layer response at a few 
percent level, since showering particles pass through alternately different 
material budget before its signal is sampled, namely W-Al-Si-PCB for the odd 
layers and W-PCB-Si-Al for the even layers.

\section{FIRST TESTBEAM RESULTS}


A structure of 10 layers with in total 60 wafers and 2160 channels was
calibrated in a cosmic testbench at Ecole Polytechnique in Paris, France.
About $1 \cdot 10^6$ cosmic triggers were recorded during December 2004.
These data allowed the calibration constants of these channels to be 
determined to 1\% level of accuracy which is sufficient for the current
phase of studies. 

\begin{figure*}[tbp]
\centering
\begin{tabular}{ccc}
      \includegraphics[width=55mm,height=53mm]{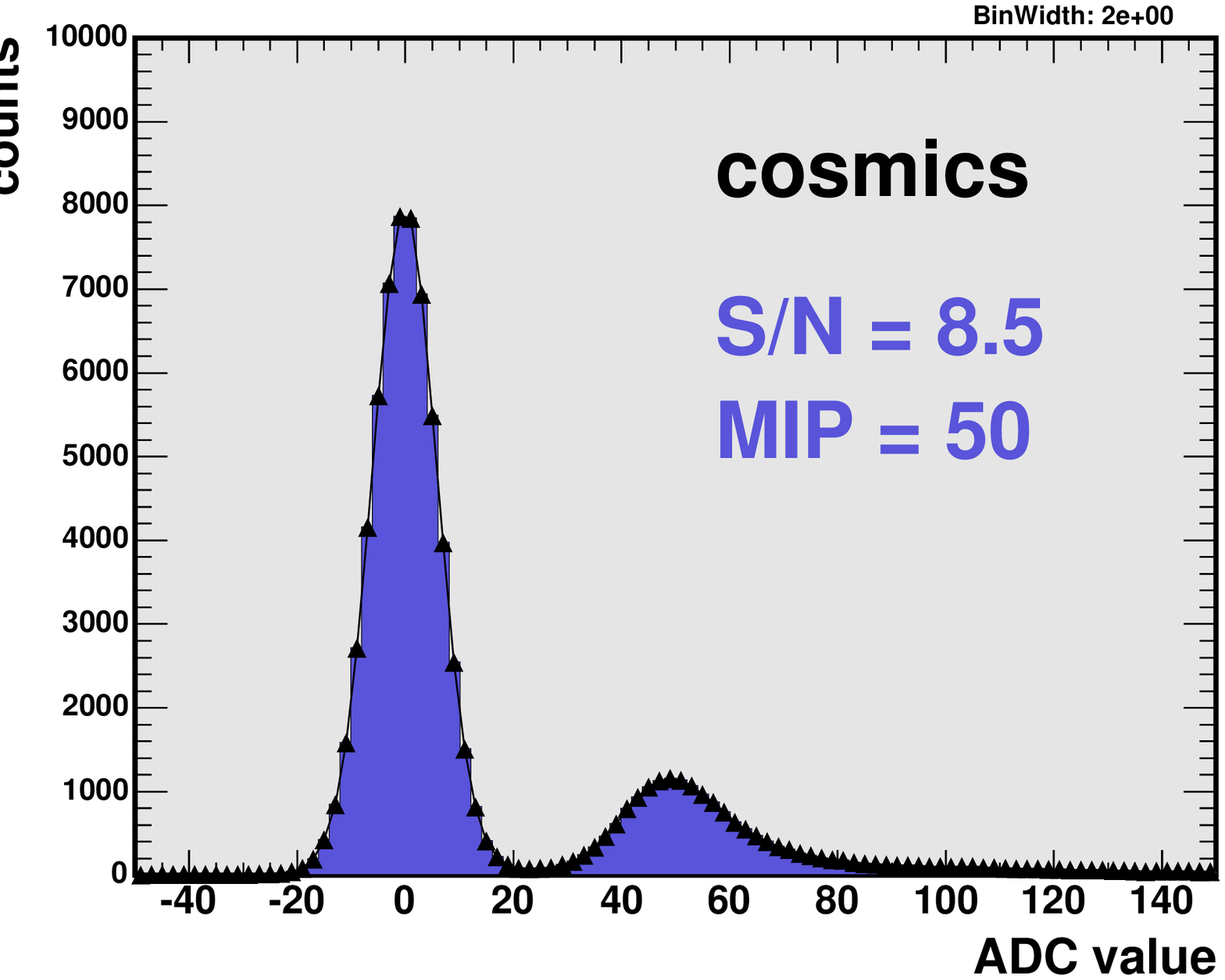}  &
      \includegraphics[width=55mm]{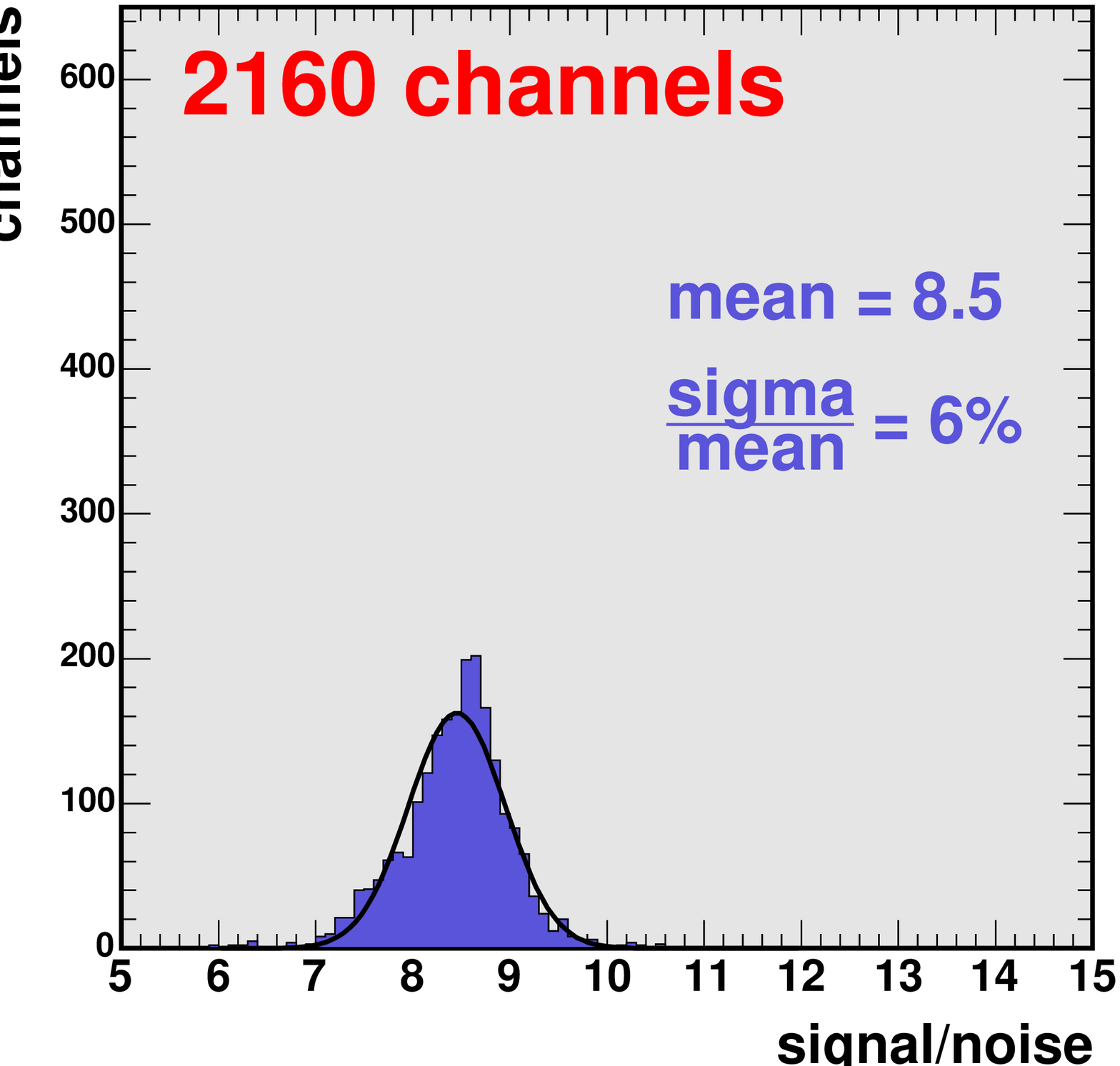}   & 
      \includegraphics[width=55mm]{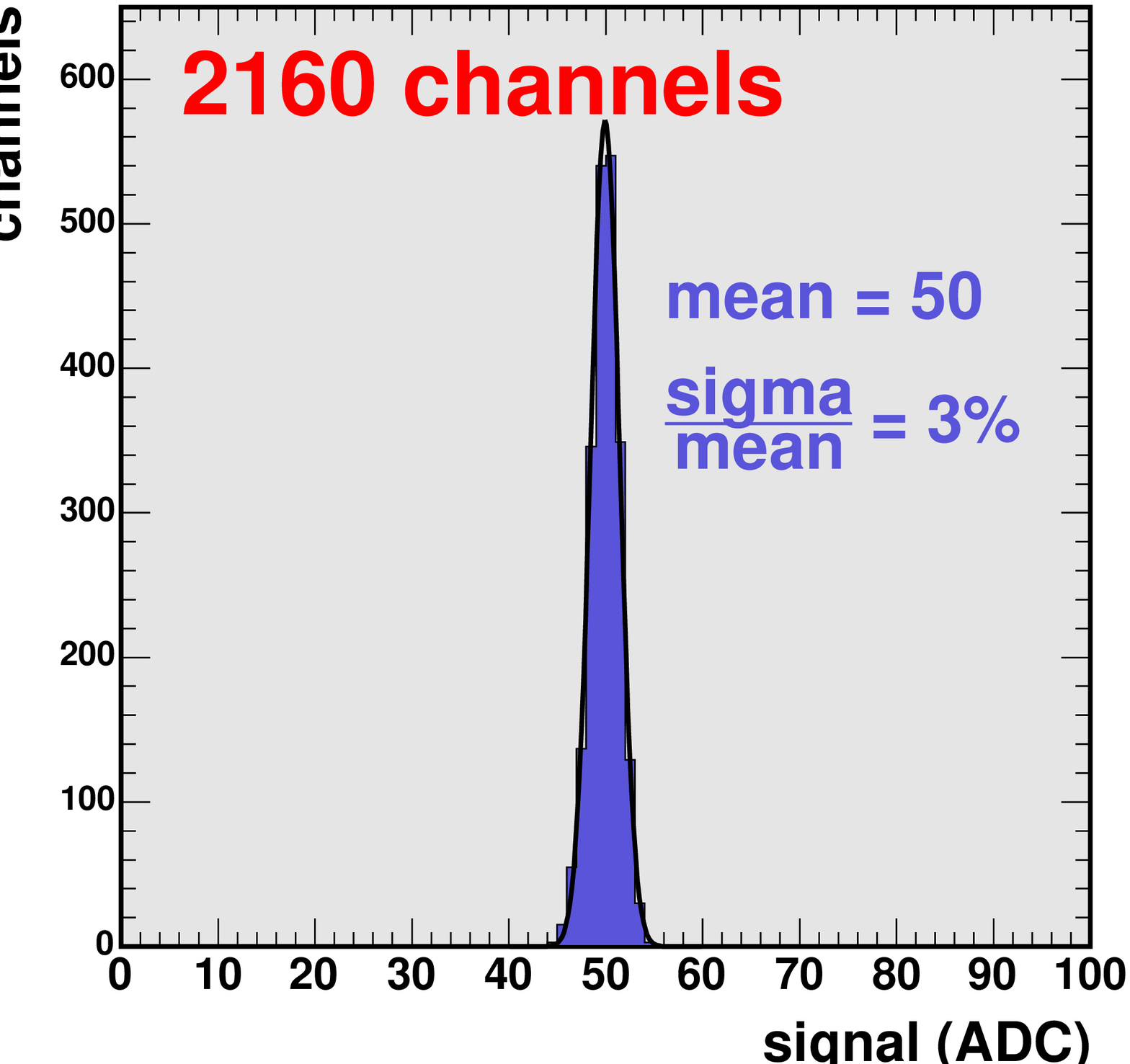}   \\		
      {\bf (a)}&
		{\bf (b)}&
		{\bf (c)}\\
   \end{tabular}
\caption{A typical channel with pedestal and mip signal with Gaussian 
and Landau shape, respectively (a). Signal over noise ratio distribution, 
(b), and mip peak distribution, (c), for the 2160 channels.} 
\label{fig:CosmicsCalibration}
\end{figure*}

A typical channel is shown in Figure~\ref{fig:CosmicsCalibration}(a), 
the pedestal and the mip signal are clearly separated and have Gaussian 
and Landau shape, respectively, as expected.
A signal over noise ratio of about 8.5 was observed with a variation 
among the channels of about 6\%, as can be seen in 
Figure~\ref{fig:CosmicsCalibration}(b). 
Figure~\ref{fig:CosmicsCalibration}(c) shows the distribution of the mip peak
for the channels. One should note how narrow the distribution is,  
with a spread, as expressed by the sigma/mean, of 3\%. This intercalibration 
spread is coming from the variation of the wafer thickness 
which is 3\% as given by the wafer suppliers. 
Construction and material inhomogeneity contribute to the constant term 
of energy resolution and is a common source of performance degradation.
To be able to determine and to correct for these inhomogeneities effects 
due to production tolerance is one of the key advantages of Si being used 
for calorimetric detectors. Since Si show relatively stable properties 
along time and with respect to temperature variation
less frequent monitoring and calibration dedicated effort will be required 
for a Si based calorimeter. 

A so-called ``30\%'' equipped Si/W prototype was shipped to DESY in Hamburg, 
Germany, and was tested with electrons during January-February 2005. 
The prototype consists of 14 W layers, the first 10 are 1.4 mm thick each and 
the last 4 at 2.8 mm, interleaved with 18~$\times$~12 matrix of active Si 
cells 1~$\times$~1~cm$^2$ each. In total the detector has 3024 channels and 
measures about 7.2~$X_0$ at normal incidence.

\begin{figure*}[tbp]
\centering
\includegraphics[width=115mm,height=55mm]{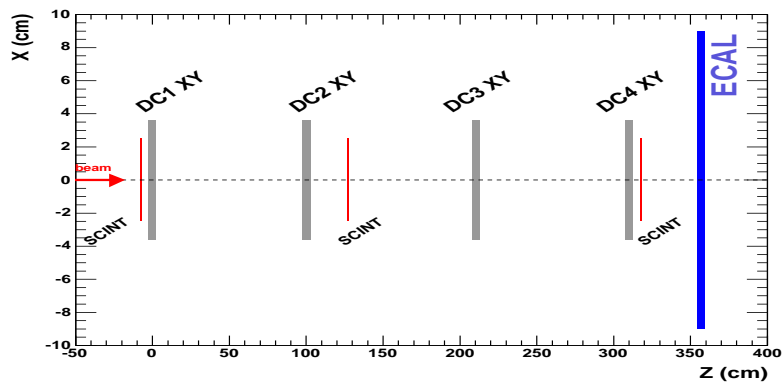}
\caption{Testbeam layout at DESY, it consists of the prototype 
detector, three scintillation counters 
to provide the trigger signal and four drift chambers to perform 
the tracking of incoming particles (the drift chambers and their installation
are courtesy of Tsukuba University and Kobe University, Japan).} 
\label{fig:Layout}
\end{figure*}

The testbeam layout is shown in Figure~\ref{fig:Layout}. Three scintillation 
counters are used to provide the trigger signal and there are also four drift 
chambers which perform the tracking of the incoming particles. The calorimeter
protoype was put on top of a moving table which allowed horizontal and 
vertical displacement with respect to beam.
Data at several configurations of position $\times$ energy $\times$ angle
were taken. We performed position scans at center - edge - corner 
of wafers, with electron beam mainly at 1, 2, 3~GeV, and with some runs 
at 4, 5 and 6 GeV, and with detector tilted with respect to the beam direction 
at $0^{\circ}$, $10^{\circ}$, $20^{\circ}$ and $30^{\circ}$.
In total about $25 \cdot 10^6$ trigger events were recorded. 
Some preliminary results are reported in the following.

\begin{figure*}[tbp]
\centering
\hspace*{40pt}\includegraphics[width=155mm]{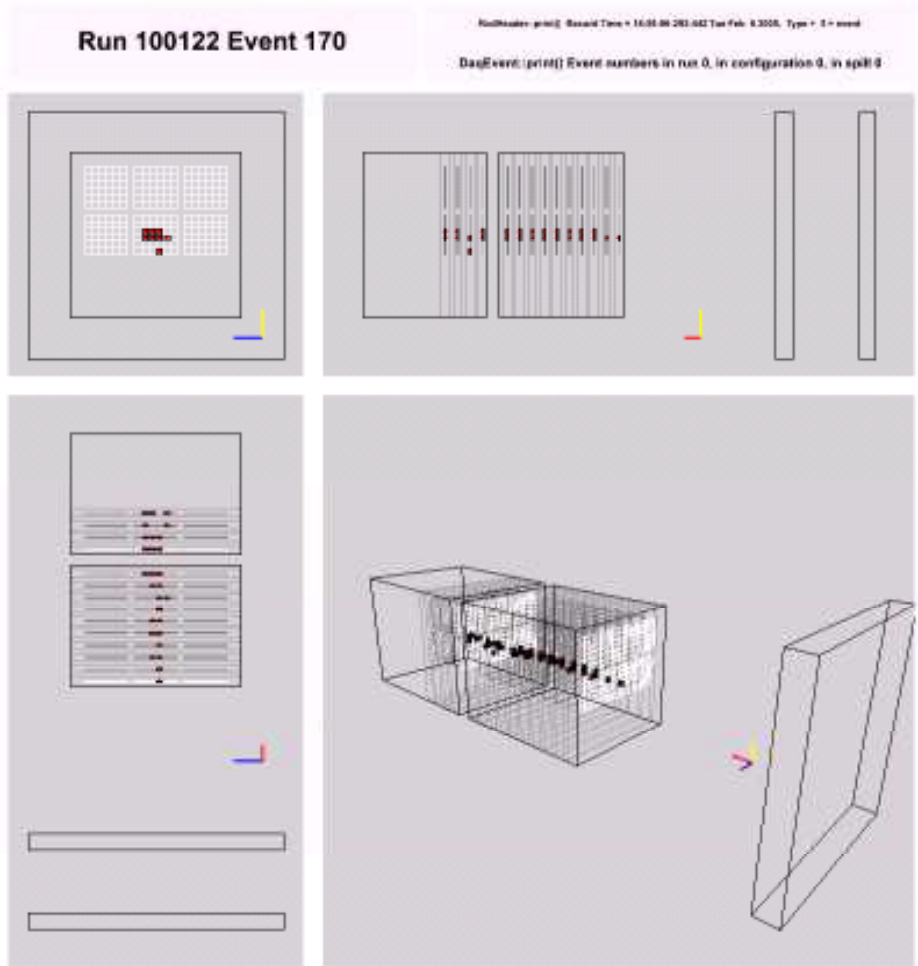}
\caption{A typical 1~GeV $e^-$ shower, the detector cells with signal 
above 0.5 mip threshold are displayed.} 
\label{fig:EventDisplay}
\end{figure*}

A typical event is displayed in Figure~\ref{fig:EventDisplay}
where the cells hit with signal above a threshold of 0.5 mip are shown.
This illustrates clearly the concept of ``tracking calorimetry'', 
the high granularity of the calorimeter allows to record the development 
of the shower along both transverse and longitudinal direction
and to reconstruct it in a tracking manner, {\em i.e.} both calorimetric and 
tracking information are provided. 

\begin{figure*}[tbp]
\centering
\hspace*{40pt}\includegraphics[width=80mm, height=60mm]{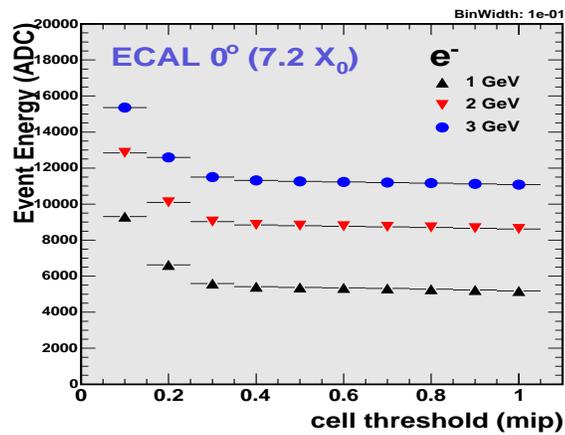}
\caption{Detector response to incident electrons as a function of the cell 
threshold applied.} 
\label{fig:ResponseVsThreshold}
\end{figure*}

The total signal recorded as a function of the cell threshold applied is 
shown in Figure~\ref{fig:ResponseVsThreshold}. A threshold around 
0.5~-~0.6~mip seems to be a safe limit in order to suppress noise sufficiently.
The results following are with a cell threshold of 0.5 mip being applied.

\begin{figure*}[tbp]
\centering
   \begin{tabular}{cc}
      \includegraphics[width=55mm]{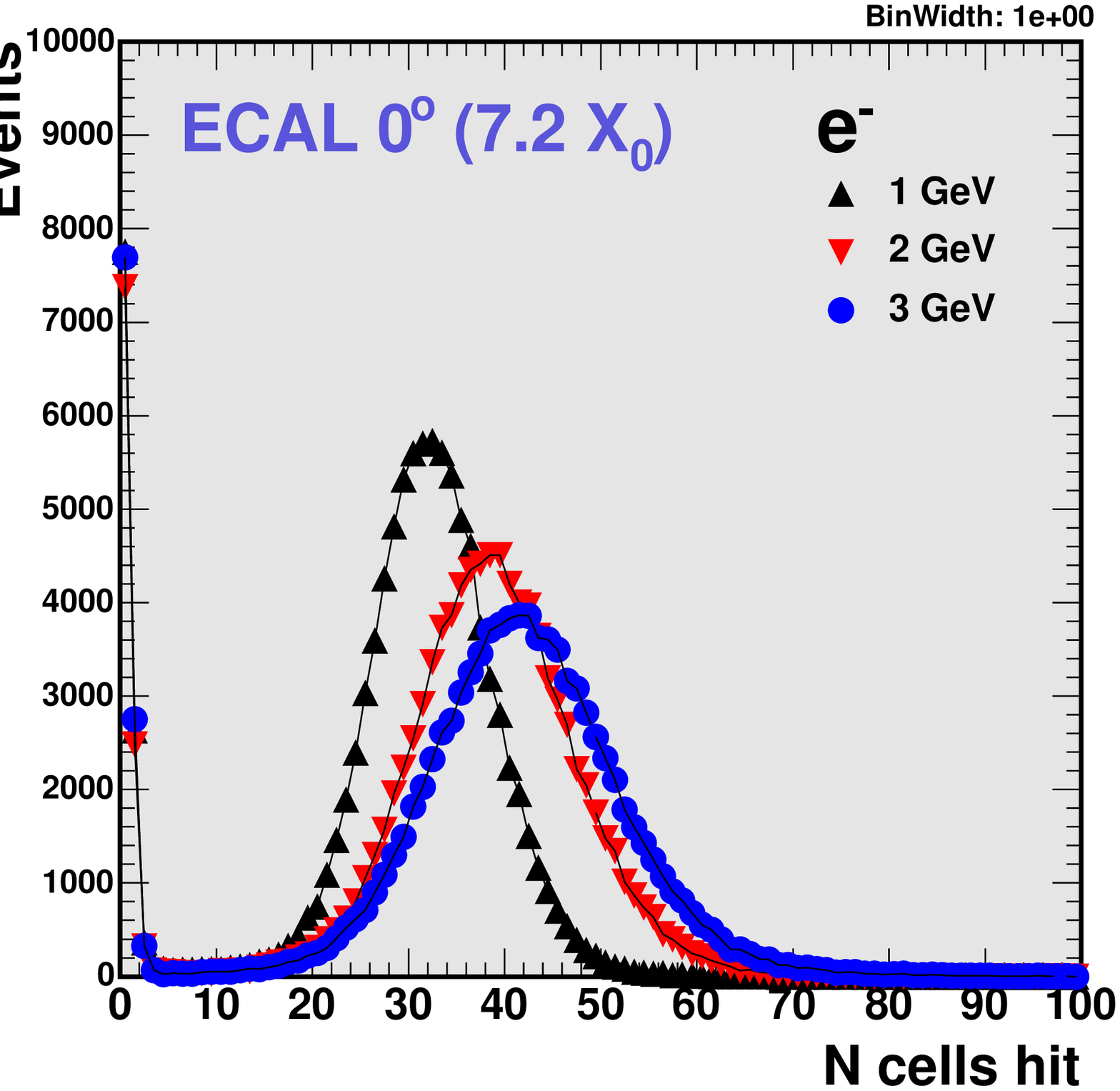}   &
      \includegraphics[width=55mm]{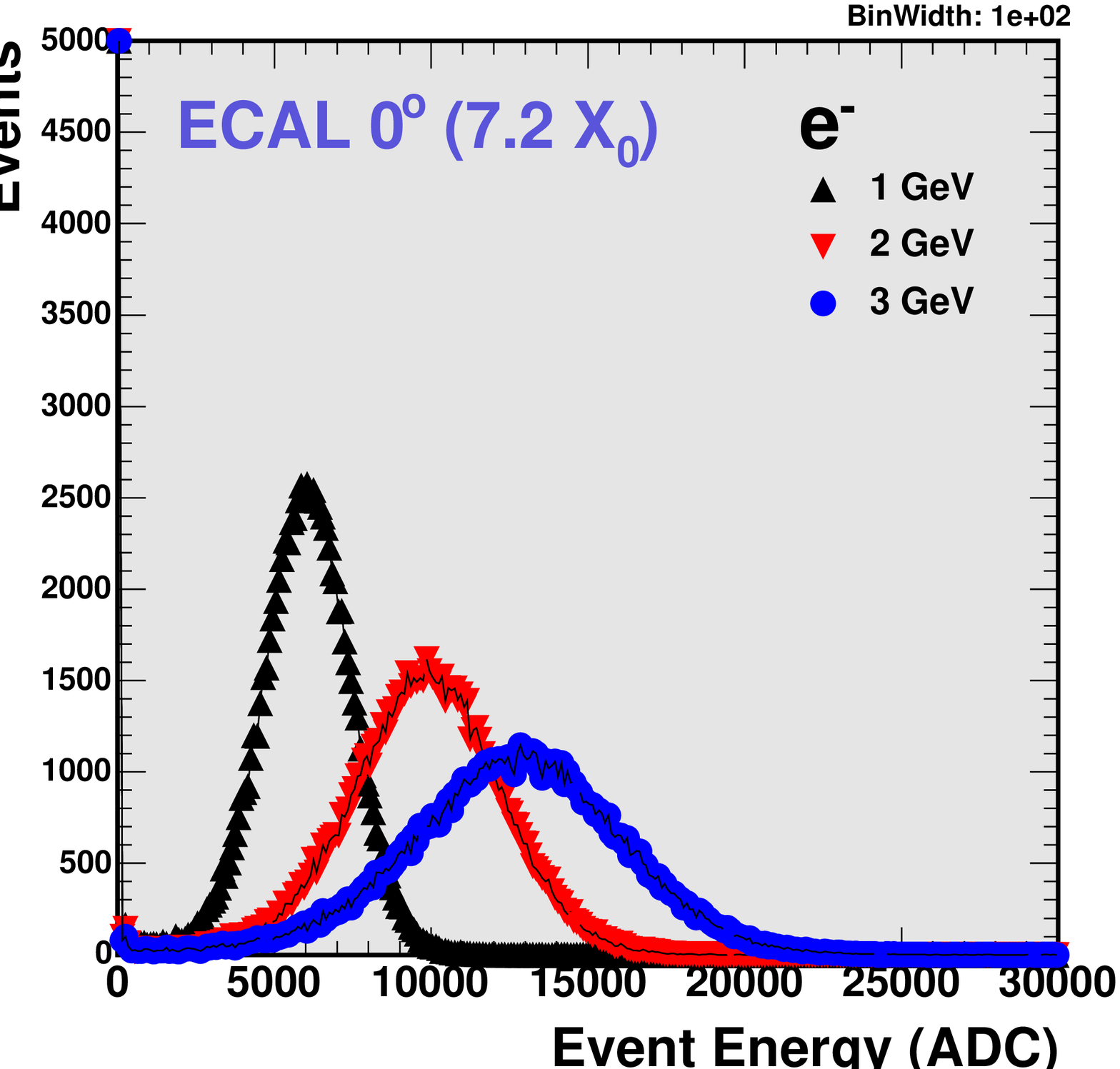}   \\
      \includegraphics[width=55mm]{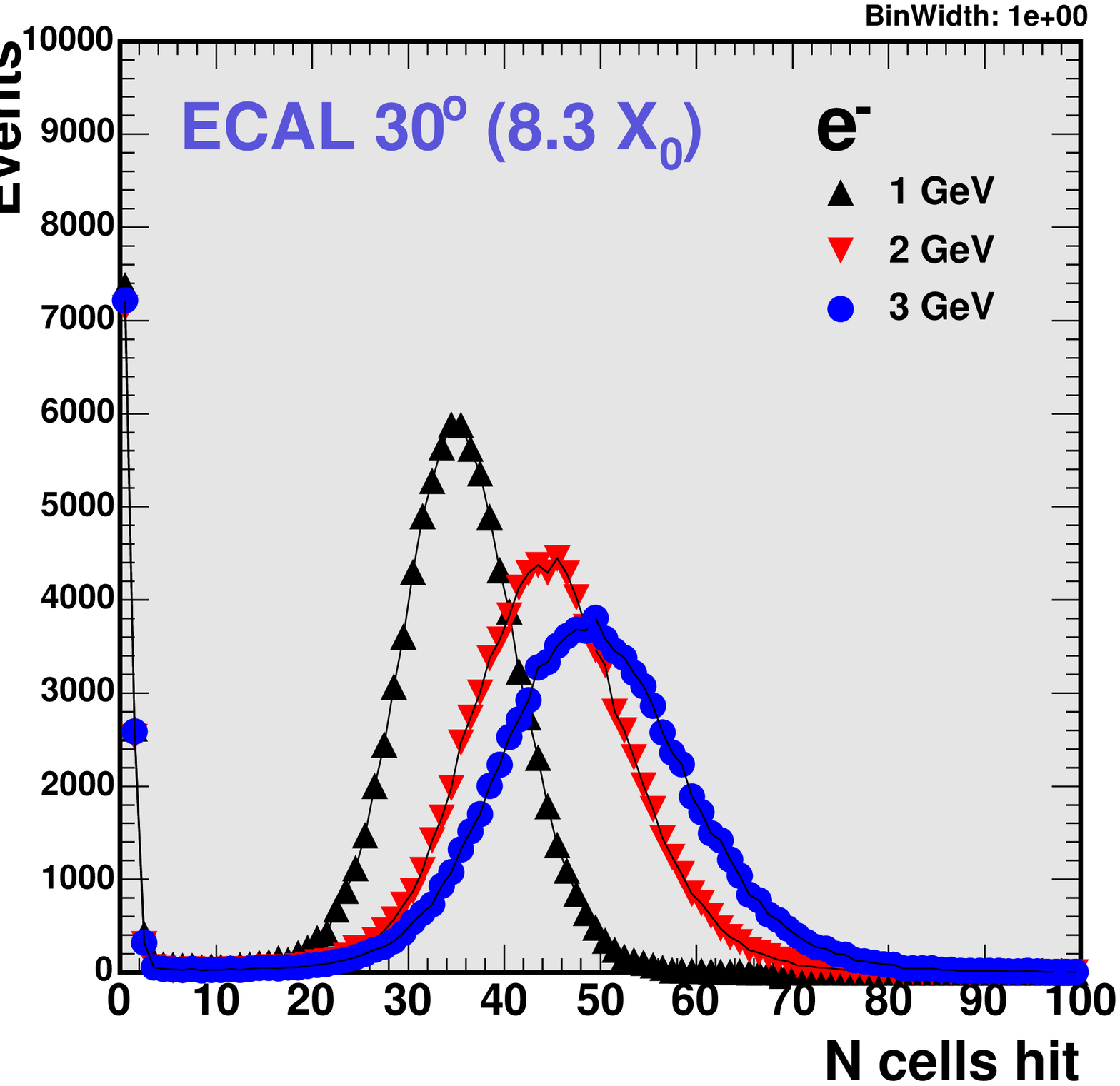}   &
      \includegraphics[width=55mm]{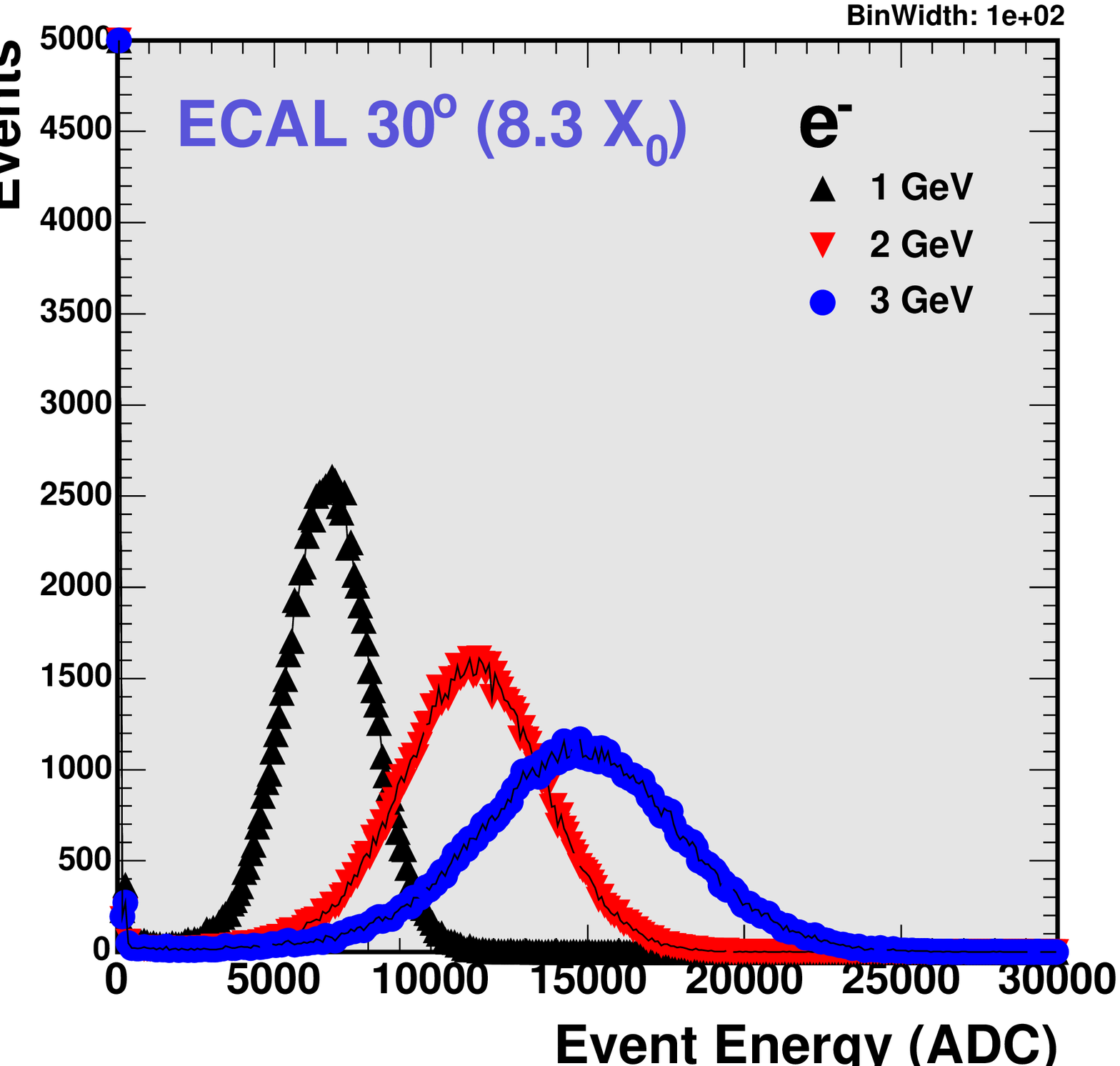}   \\
   \end{tabular}
\caption{Response to electrons, in terms of total number of cells hit 
(left column) and energy deposited (right column) for detector being 
tilted at $0^{\circ}$ (upper row) or $30^{\circ}$ (lower row) with respect 
to beam direction.} 
\label{fig:Response}
\end{figure*}

\begin{figure*}[tbp]
\centering
   \begin{tabular}{cc}
      \includegraphics[width=55mm]{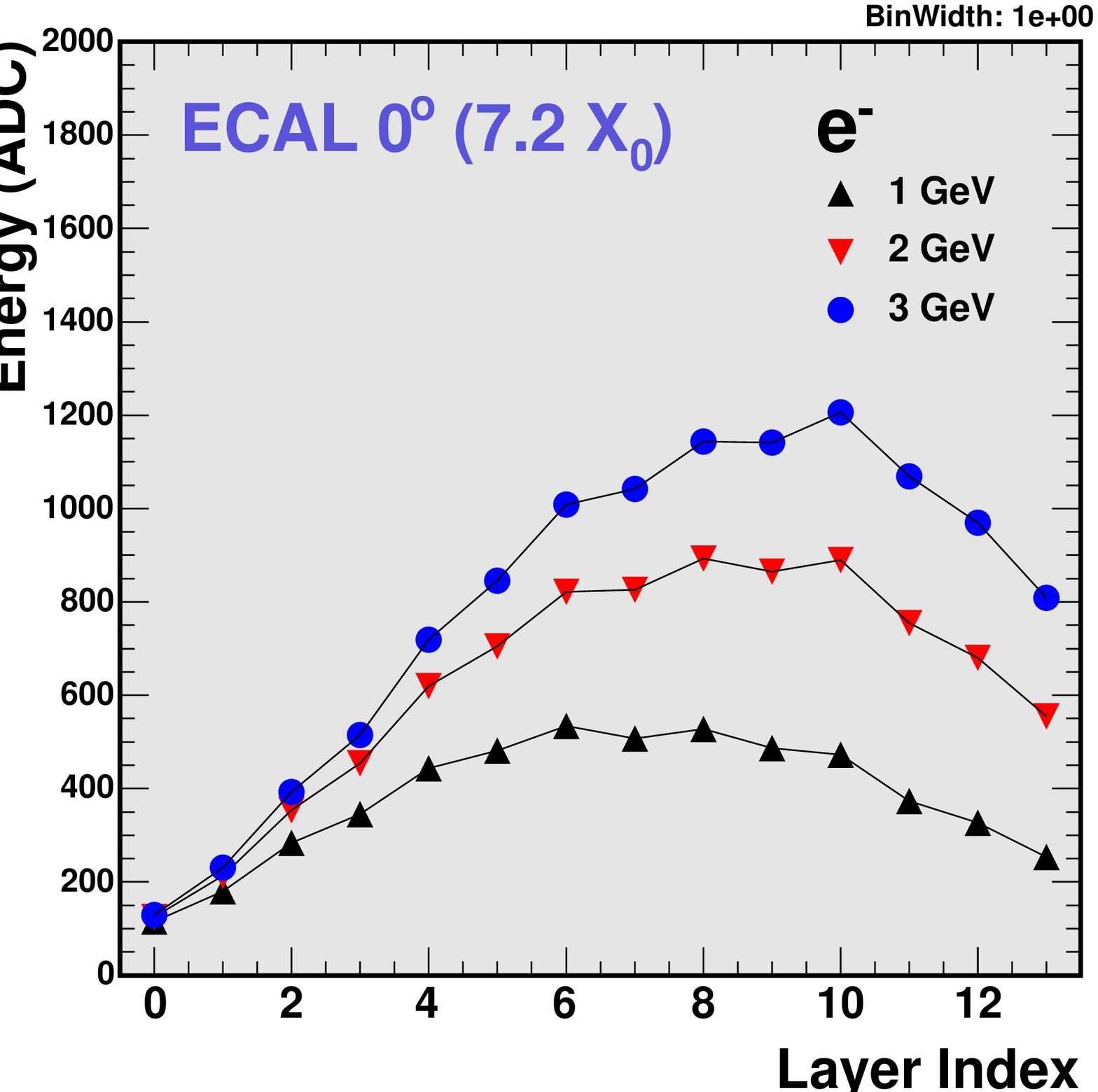}   &
      \includegraphics[width=55mm]{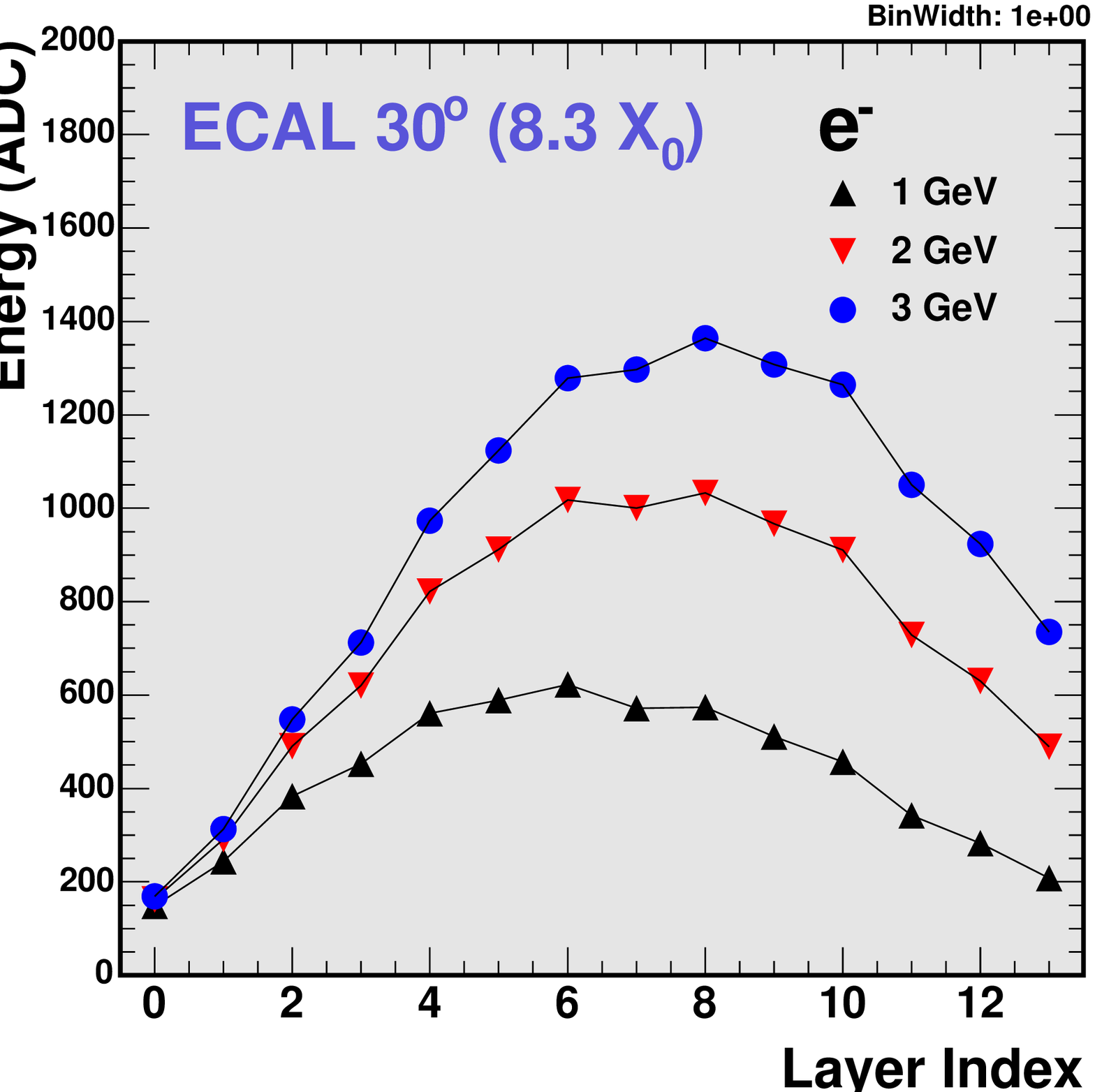}   \\
   \end{tabular}
\caption{Longitudinal development of electron showers. Average response per 
layer and for detector being tilted at $0^{\circ}$ (left) or 
$30^{\circ}$ (right) with respect to beam direction.} 
\label{fig:LongitudinalDevelopment}
\end{figure*}

Figure~\ref{fig:Response} shows the detector response to incident electrons 
with energy of 1, 2 and 3~GeV in terms of total number of cells above 0.5~mip 
threshold (left column) or total energy deposited (right column). The upper 
and lower row of plots correspond
to prototype being at $0^{\circ}$ and $30^{\circ}$ angle with respect to 
beam direction, respectively. 
The prototype is not long enough to contain the showers and therefore its 
response shows poor linearity. Better containment is achieved at 
$30^{\circ}$ as can be seen also in Figure~\ref{fig:LongitudinalDevelopment}
which depicts the longitudinal development of the showers. The odd/even 
layer response asymmetry due to construction is observed.

\begin{figure*}[tbp]
\centering
\begin{tabular}{ccc}
      \includegraphics[width=55mm]{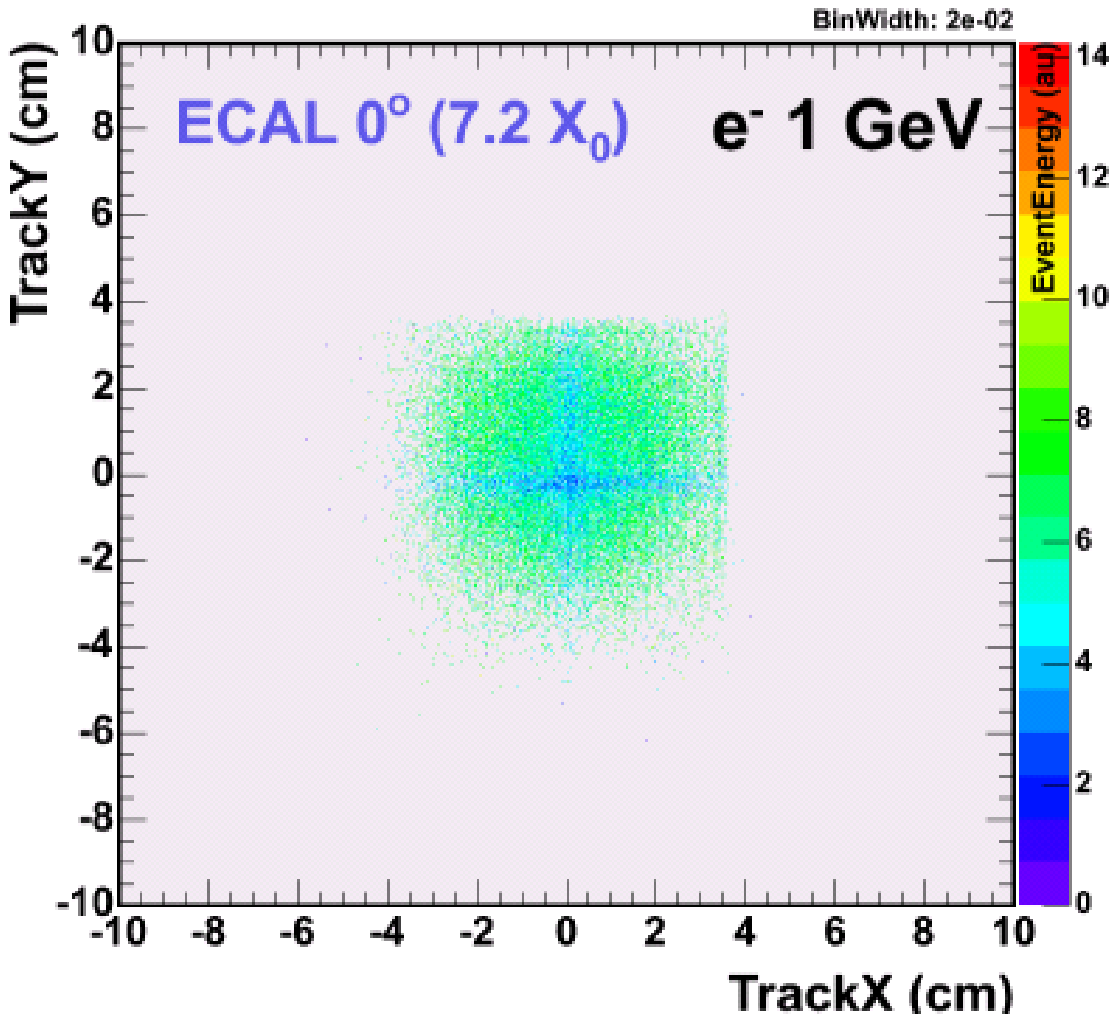}   & 
      \hspace{15pt}
      \includegraphics[width=55mm]{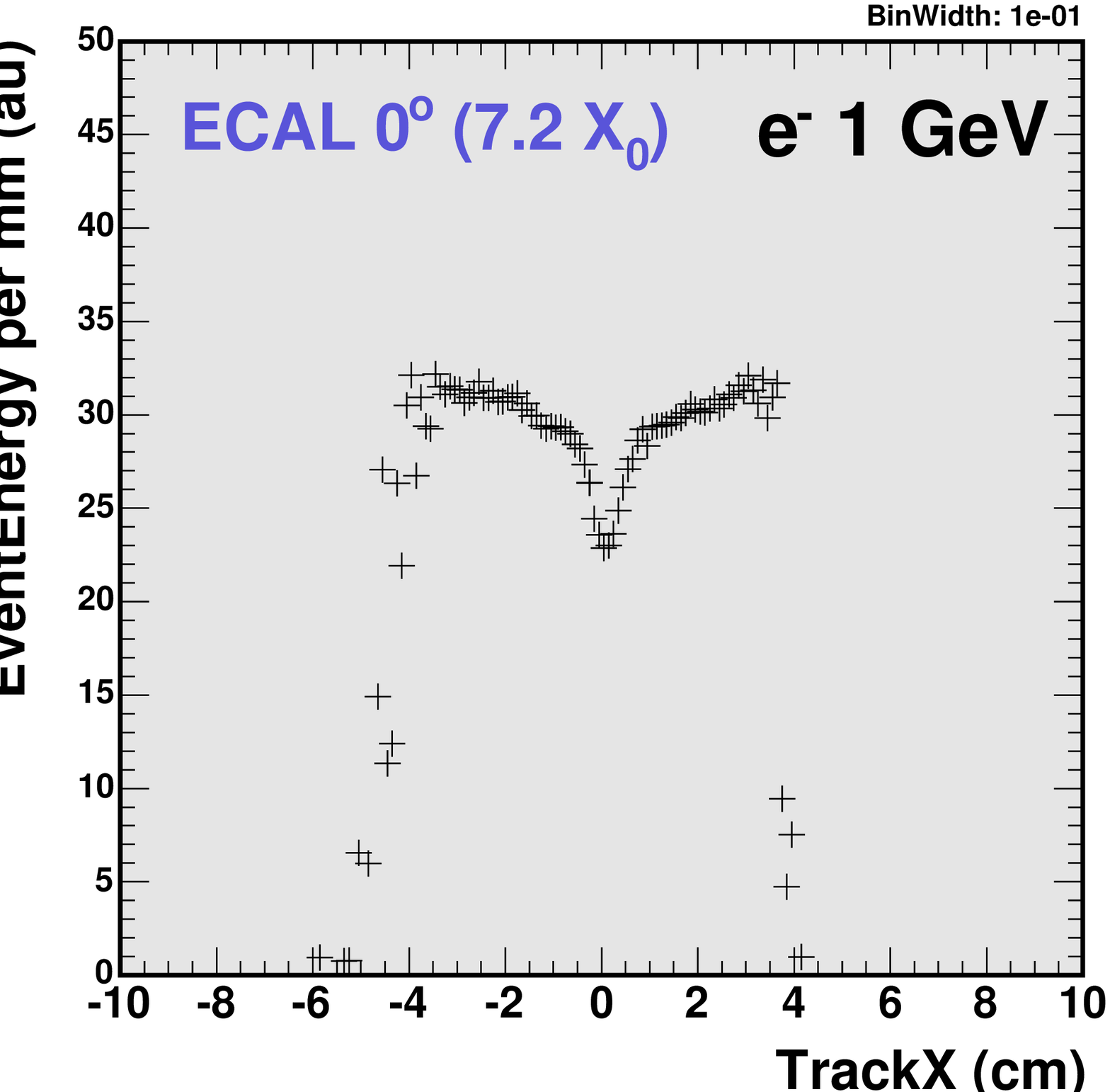}   &
      \includegraphics[width=55mm]{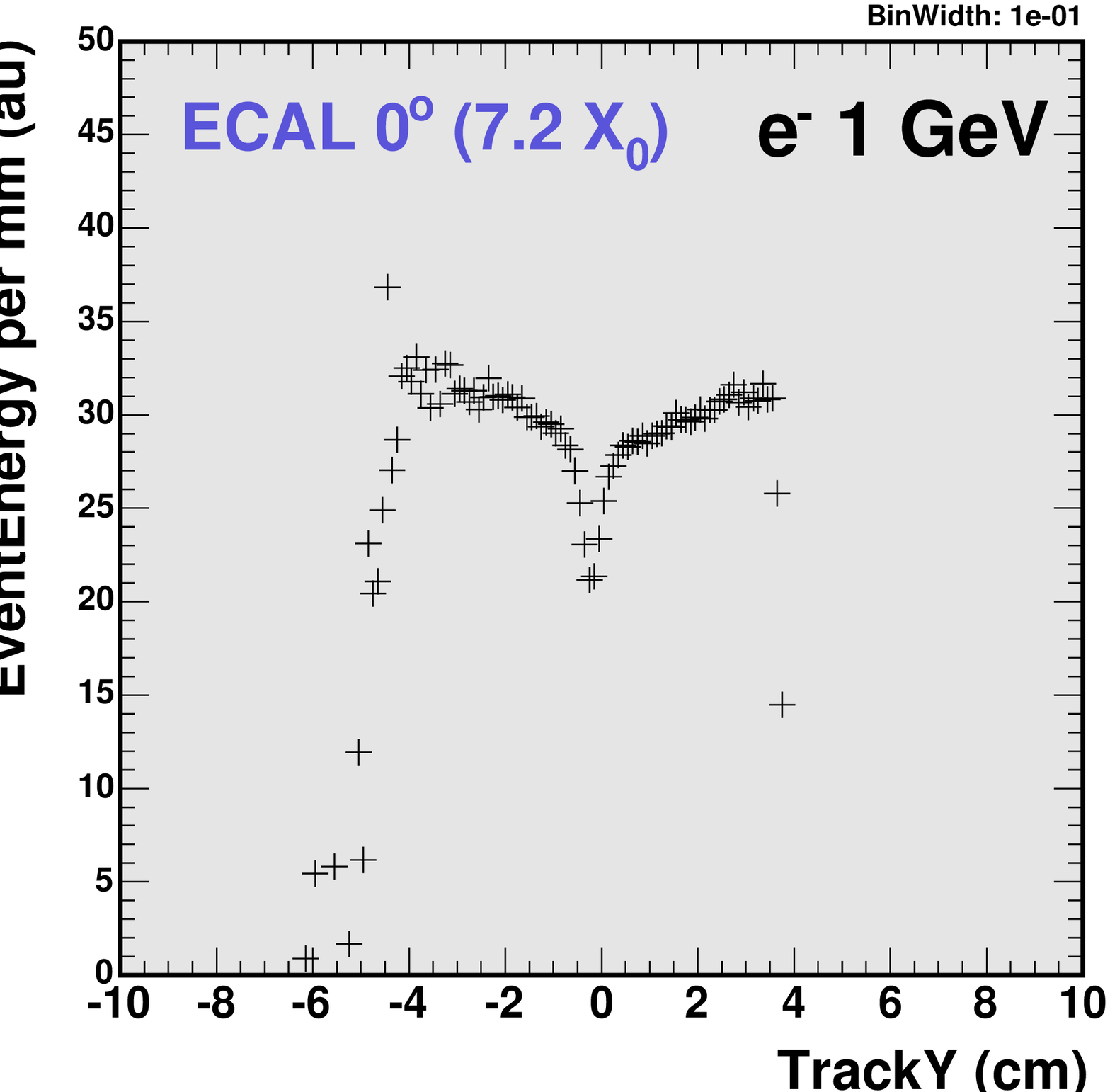}   \\     
      {\bf (a)}&
      {\bf (b)}&
      {\bf (c)}\\
\end{tabular}
\caption{Position scan at the corners of four neighboring wafers. 
Detector response, expressed in color coded scale arbitrary units, 
versus beam impact point (a), projection along the horizontal direction (b), 
projection along the vertical direction (c).} 
\label{fig:PositionScanCorner}
\end{figure*}

Several position scans along the wafers and their borders
were performed to investigate the response homogeneity.
The plots at Figure~\ref{fig:PositionScanCorner} illustrate 
the results from a typical scan at the corners of four neighboring wafers.
Figure~\ref{fig:PositionScanCorner}(a) is the scatter plot of the 
event energy recorded, expressed in color coded scale, versus the 
impact point coordinates of the electron beam.
Each wafer has a border of about 1~mm of non-active zone. 
The drop of signal along the borders is clearly seen. 
Figures~\ref{fig:PositionScanCorner}(b) and (c) are the 
corresponding projections along the horizontal and the vertical direction.
The alternate layers of the detector are staggered horizontally, but not 
vertically. Therefore, as observed, the dip is shallower and wider in the 
former case, and deeper and  narrower in the latter one. 

Similar studies are under way for general debugging and understanding 
of the system before the next testbeam phase. 

\section{SUMMARY}

We discussed the CALICE Si/W ECAL prototype and some first results from a 
recent testbeam. A prototype with 14 layers and depth of 7.2~$X_0$ 
at normal incidence, having in total 3024 channels of 1~$\times$~1~cm$^2$ 
Si pads, was tested at DESY with $e^-$ beam. The testbeam progressed very 
smoothly and a lot of data were collected that will allow us to perform a 
thorough debugging of the system. Analysis studies are in progress  
in order to understand the detector before the next round of testbeams. 
The next testbeam is planned for summer 2005 with the prototype structure 
equipped with more layers and channels.

\begin{acknowledgments}

On behalf of the ECAL group the author wish to thank 
the CALICE-HCAL group at DESY for their support and contribution 
to the success of the testbeam.

\end{acknowledgments}



\end{document}